\documentclass[aps,prl,twocolumn,floatfix,showpacs]{revtex4}

\usepackage{graphics}
\usepackage{graphicx}
\usepackage{amssymb}
\usepackage{amsfonts}
\usepackage{amsmath}
\usepackage{bm}

\begin{document}

\title{Spontaneous Spin Polarization  in Quantum Wires}
\author{A. D. Klironomos}
\author{J. S. Meyer}
\thanks{On leave from The Ohio State University, Columbus, OH 43210-1117.}
\author{K. A. Matveev}
\thanks{On leave from Duke University, Durham, NC 27708-0305.}
\affiliation{Materials Science Division, Argonne National Laboratory, Argonne, Illinois 60439, USA}
\date{\today}

\begin{abstract}

  A number of recent experiments report spin polarization in quantum wires
  in the absence of magnetic fields. These observations are in apparent
  contradiction with the Lieb-Mattis theorem, which forbids spontaneous
  spin polarization in one dimension. We show that sufficiently strong
  interactions between electrons induce deviations from the strictly
  one-dimensional geometry and indeed give rise to a ferromagnetic ground
  state in a certain range of electron densities.

\end{abstract}


\maketitle

Quantum wires are quasi-one-dimensional structures which, although
conceptually simple, display extremely rich physics that defies
conventional intuition developed for two- and three-dimensional
conductors. The study of transport properties of quantum wires has offered
a direct glimpse into the quantum world through the quantization of
conductance in integer multiples of $G_0=2e^2/h$ \cite{Wees}. Recently,
one of the most exotic implications of one-dimensionality---the existence
of separate spin and charge excitations---has been demonstrated
experimentally \cite{Auslaender}.

In a number of recent experiments on quantum wires, deviations from perfect
conductance quantization have been observed
\cite{Thomas,Kane,Thomas_2,Reilly,Morimoto,Kristensen,Cronenwett}.
Most commonly the experimental findings have been interpreted as
indication of spontaneous spin polarization
\cite{Thomas,Kane,Thomas_2,Reilly,Morimoto,Kristensen}. However,
for a strictly one-dimensional system this
possibility is explicitly forbidden by a theorem due to E. Lieb
and D. Mattis \cite{Lieb}, based on very general mathematical
properties of the Schr\"odinger equation describing these interacting
electronic systems. Although a number of interpretations of the conductance
anomalies that do not rely on the idea of spin polarization have been
proposed \cite{Cronenwett,Tokura,Meir,Matveev}, the experiments do
raise a fundamental question: {\it Can the ground state of the
electron system in a quantum wire be ferromagnetic?}

The only way to circumvent the Lieb-Mattis theorem is to
recognize that realistic quantum wires are not in essence
one-dimensional devices. Attempts in that direction have been made
\cite{Spivak}, requiring, however, a fully two-dimensional structure
as a starting point. By contrast, we start with the conventional model
of an electron gas in a quantum wire and show that strong Coulomb
interactions both cause deviations from one-dimensionality and bring
about a ferromagnetic ground state.

Typical experiments are done with quantum wires that are formed at the
interfaces of GaAs/AlGaAs heterostructures. A voltage applied to metal
gates provides a confining potential in the directions transverse to the
wire and, in addition, allows one to tune the electron density in the
wire. While conductance plateaus at integer multiples of $G_0$ are
observed in the high density regime, a drop in conductance commonly
attributed \cite{Thomas,Kane,Thomas_2,Reilly,Morimoto,Kristensen} to spin
polarization has been observed
\cite{Thomas,Kane,Thomas_2,Reilly,Morimoto,Kristensen,Cronenwett} in the
region of gate voltages where the electron density is very low.

As the density $n$ of electrons is lowered, Coulomb interactions
become more important, and at $n\ll a_B^{-1}$ they dominate over the
kinetic energy.  (Here $a_{B}=\hbar^{2}\epsilon/me^{2}$ is the Bohr radius
in the material, $\epsilon$ is its dielectric constant, and $m$ is the
effective electron mass; $a_{B}\!\approx\!100${\AA} in GaAs.)  In this
limit the electrons can be viewed as classical particles.  In order to
minimize their mutual Coulomb repulsion, electrons occupy equidistant
positions along the wire, forming a structure with short-range crystalline
order---the so-called Wigner crystal.  Upon increasing the density, the
inter-electron distance diminishes, and the resulting stronger electron
repulsion eventually overcomes the confining potential, transforming the
classical one-dimensional Wigner crystal into a staggered or zig-zag chain
\cite{Piacente}. Typical structures for different densities are shown in
Fig.~1.

\begin{figure}[!b]
\vspace*{-0.6cm}
\centerline{\includegraphics[height=5.8cm,clip]{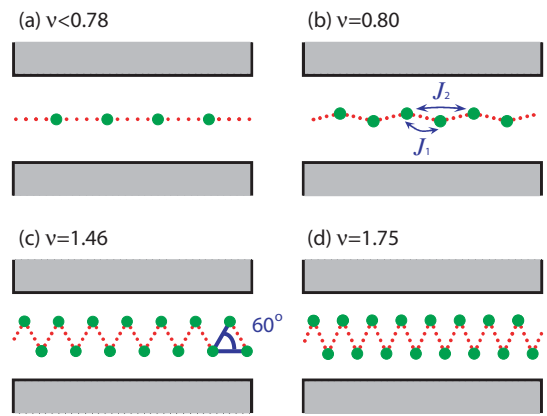}}
\caption{Wigner crystal of electrons in a quantum
wire defined by gates (shaded). The structure is determined by the
parameter $\nu$ proportional to electron density (see text). As
density grows, the one-dimensional crystal (a) gives way to a
zig-zag chain (b-d). The arrows in (b) illustrate the
nearest neighbor $(J_{1})$ and next-nearest neighbor $(J_{2})$
exchange processes.}
\label{fig_1}
\end{figure}

Quantum-mechanically, spin-spin interactions in the Wigner crystal arise
due to exchange processes, in which two electrons switch positions by
tunneling through the potential barrier that separates them. The barrier
is created by the two exchanging particles as well as all other electrons
in the wire.  Originating in tunneling, the exchange energy associated
with such processes falls off exponentially with the distance between the
electrons.  As a result, only the nearest-neighbor exchange is relevant in
a one-dimensional crystal.  The corresponding exchange constant is
positive, leading to an antiferromagnetic ground state in accordance with
the Lieb-Mattis theorem \cite{Lieb}.

A very different situation arises when one considers the most trivial
deviation from the one-dimensional crystal, namely the zig-zag chain
introduced above. For that structure, depending on the distance
between the two rows which varies as a function of density, the
distance between next-nearest neighbors may be equal to or even
smaller than the distance between nearest neighbors, as illustrated in
Fig.~1(c,d). Accordingly, the next-nearest neighbor exchange constant
$J_{2}$ may be equal to or larger than the nearest neighbor exchange
constant $J_{1}$. The corresponding spin chain is described by the
Hamiltonian
\begin{equation}
H_{12} = \sum_j\left(J_1{\bf S}_j{\bf S}_{j+1}+J_2{\bf S}_j{\bf S}_{j+2}\right).
\label{eq-H_12}
\end{equation}
The competition between the two exchanges causes frustration of the
antiferromagnetic spin order and eventually leads to a gapped dimerized
ground state at $J_{2}>0.24J_{1}$, \cite{Majumdar,Haldane,Eggert}. In
addition, drawing intuition from studies of the two-dimensional Wigner
crystal, one realizes that in this geometry \emph{ring-exchange}
processes, in which three or more particles exchange positions in a cyclic
fashion, have to be considered.

It has been established that, due to symmetry properties of the ground
state wave functions, ring exchanges of an even number of fermions
favor antiferromagnetism, while those of an odd number of fermions
favor ferromagnetism \cite{Thouless}. In a zig-zag chain, the Hamiltonian reads
\begin{eqnarray}
H&\!\!=\!\!&\frac12\sum_j\Big(
J_1 P_{j\,j\!+\!1}+J_2 P_{j\,j\!+\!2}-J_3(P_{j\,j\!+\!1\,j\!+\!2}
+P_{j\!+\!2\,j\!+\!1\,j})\nonumber\\
&&\quad\quad\quad + J_4(P_{j\,j\!+\!1\,j\!+\!3\,j\!+\!2}+P_{j\!+\!2\,j\!+\!3\,j\!+\!1\,j})
-\dots\Big),
\label{eq:H}
\end{eqnarray}
where the exchange constants are defined such that all $J_l>0$ and
only the dominant $l$-particle exchanges are shown. Here,
$P_{j_1\dots j_l}$ denotes the cyclic permutation operator of $l$ spins. A more
familiar form of the Hamiltonian in terms of spin operators is obtained using
that $P_{ij}=\frac{1}{2} +2{\bf S}_i{\bf S}_j$ and
$P_{j_1\dots j_l}=P_{j_1j_2}P_{j_2j_3}\dots P_{j_{l-1}j_l}$ \cite{Thouless}. In
particular, the two-spin exchanges reduce to Eq.~(\ref{eq-H_12}).

The simplest ring exchange involves three particles and is therefore
ferromagnetic. Extensive studies of the two-dimensional Wigner crystal have
shown that, at low densities (or strong interactions), the three-particle
ring exchange dominates over the two-particle exchange. As a result,
the two-dimensional Wigner crystal becomes ferromagnetic at
sufficiently strong interactions \cite{Roger,Bernu}. Since the
electrons in a two-dimensional Wigner crystal form a triangular
lattice, by analogy, one should expect a similar effect in the
zig-zag chain at densities where the electrons form approximately
equilateral triangles, Fig.~1(c). In order to verify this scenario,
we have to identify the electron configuration that is stable at a
given density and subsequently find the corresponding exchange
energies.

Specifically, we consider a quantum wire with a parabolic confining
potential $V_{\rm conf}(y)=m\Omega^{2}y^{2}/2$, where $\Omega$ is the
frequency of harmonic oscillations in the potential $V_{\rm conf}(y)$. At
low electron density $n$ in the wire, a one-dimensional Wigner crystal
is formed, Fig.~1(a). As the density grows, however, the Coulomb
interaction energy becomes comparable to the confining potential, leading
to the formation of a zig-zag chain, as depicted in Fig.~1(b-d). This
transition happens when distances between electrons are of the order of
the characteristic length scale
$r_{0}=\left(2e^2/\epsilon m\Omega^2\right)^{1/3}$, such that
$V_{\rm conf}(r_{0})= V_{\rm int}(r_{0})$, where $V_{\rm
  int}(r)=e^{2}/\epsilon r$ is the Coulomb interaction energy. It is
convenient for the following discussion to introduce a dimensionless
density $\nu=nr_{0}$. Minimization of the energy with respect to the
electron configuration \cite{Piacente} reveals that a one-dimensional
crystal is stable for densities $\nu<0.78$, whereas a zig-zag chain forms
at intermediate densities $0.78<\nu<1.75$. (At higher densities, the
zig-zag chain gives way to structures with larger numbers of rows
\cite{Piacente}.) The distance between rows grows with density, and the
equilateral configuration is achieved at $\nu\approx 1.46$, well within
the region where the zig-zag chain is stable. Therefore, there are strong
indications that the ferromagnetic state may be realized. More
specifically, upon increasing the density one would expect the system to
undergo two consecutive phase transitions: first from an antiferromagnetic
to a ferromagnetic, and then to a dimer phase.  However, the latter
scenario cannot be established conclusively based solely on the
two-dimensional Wigner crystal physics. The main differences are (i) the
presence of a confining potential as opposed to the flat background in the
two-dimensional case, and even more importantly, (ii) the change of the
electron configuration with density, Fig.~1, as opposed to the ideal
triangular lattice in two dimensions. Below, we study numerically the
exchange energies for the specific configurations of the zig-zag Wigner
crystal in a parabolic confining potential.

The strength of the interactions is characterized by the parameter
\begin{equation}
r_{\Omega}=\frac{r_{0}}{a_{B}}=2\left(
\frac{me^4}{2\epsilon^2\hbar^2}\,\frac{1}{\hbar\Omega}\right)^{2/3}.
\end{equation}
For $r_{\Omega}\gg 1$, the physics of the system is dominated by strong
interactions, and a semiclassical description is applicable.  In order to
calculate the various exchange constants, we use the standard instanton
method, also employed in the study of the two-dimensional Wigner crystal
\cite{Roger,Voelker}. Within this approach, the exchange constants are
given by $J_{l}=J^{*}_{l}\exp{(-S_{l}/\hbar)}$, where $S_{l}$ is the value
of the Euclidean (imaginary time) action, evaluated along the classical
exchange path. By measuring length and time in units of $r_0$ and
$T=\sqrt2/\Omega$, respectively, the action $S[\{{\bf r}_j(\tau\}] $ is
rewritten in the form $S=\hbar\eta\sqrt{r_\Omega}$, where the functional
\begin{equation}
\eta[\{{\bf r}_j(\tau)\}]=\!\int\limits_{-\infty}^\infty\!\!\mathrm{d}\tau
     \!\left[\sum_{j}
     \left(\frac{\dot{{\bf r}}_j^2}{2}+y_j^2\right)
    +\sum_{j<i}\frac{1}{|{\bf r}_j\!-\!{\bf r}_i|}
  \right]\!\!
 \label{eq:action}
\end{equation}
is dimensionless.

Thus, we find the exchange constants in the form
$J_{l}=J^{*}_{l}\exp{(-\eta_{l}\sqrt{r_\Omega})}$, where the dimensionless
coefficients $\eta_l$ depend only on the electron configuration (cf.~Fig.~1)
or, equivalently, on density $\nu$. The instanton trajectories, and
subsequently the exponents $\eta_{l}$, are calculated for each type of
exchange by solving the equations of motion obtained from the
dimensionless action (\ref{eq:action}) numerically.

\begin{figure}[!b]
\centerline{\includegraphics[height=5.8cm,clip]{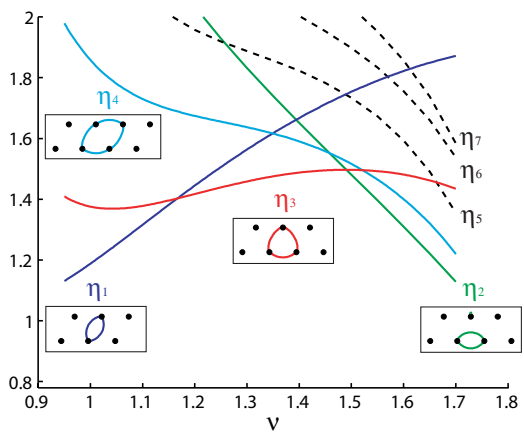}}
\caption{The exponents $\eta_{l}$ as functions of the
dimensionless density $\nu$, computed with frozen spectators. The
insets illustrate the four most important exchange processes.}
\label{fig_2}
\end{figure}

To first approximation, we neglect the motion of all ``spectators''---the
electrons in the crystal to the left and to the right of the exchanging
particles.  Figure 2 shows the calculated exponents for various exchanges
as a function of dimensionless density $\nu$. At strong interactions
($r_{\Omega}\gg1$), the exchange with the smallest value of $\eta_{l}$ is
clearly dominant, and the prefactor $J^{*}_{l}$ is of secondary importance
to our argument. The numerical calculation confirms our original
expectation: the dominant exchange constant changes from nearest neighbor
exchange $J_{1}$ to three-particle ring exchange $J_{3}$ to next-nearest
neighbor exchange $J_{2}$. More complicated ring exchanges have also been
computed. Figure 2 displays the ones with the smallest exponents, namely
the four-particle ring exchange as well as five-, six-, and seven-particle
ring exchanges (dashed lines).

\begin{figure}[!t]
\centerline{\includegraphics[height=5.8cm,clip]{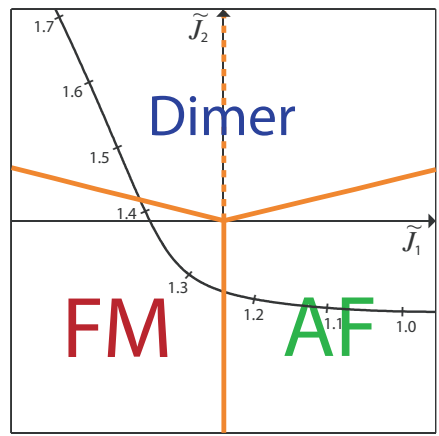}}
\caption{The phase diagram including nearest neighbor,
next-nearest neighbor, and three-particle ring exchanges. The effective
couplings $\widetilde{J}_1$ and $\widetilde{J}_2$ are defined in the
text. The solid line shows schematically the traversal of the various
phases with increasing dimensionless density $\nu$, as dictated by the
calculated exchange energies.}
\label{fig_3}
\end{figure}

If one includes only the dominant exchanges $J_1$, $J_2$, and $J_3$, the
Hamiltonian of the corresponding spin chain takes a simple form. Nearest
and next-nearest neighbor exchanges are described by Eq.~(\ref{eq-H_12}).
Furthermore, the three-particle ring exchange does not introduce a new
type of coupling, but modifies the two-particle exchange constants
\cite{Thouless}.  For a zig-zag crystal we find
\begin{equation}
H_3 = -J_3\sum_j\big(2{\bf S}_j{\bf S}_{j+1}+{\bf S}_j{\bf
  S}_{j+2}\big).
\label{eq-H3}
\end{equation}
Thus the total Hamiltonian still has the form (\ref{eq-H_12}), but with
the effective two-particle exchange constants
$\widetilde{J}_{1}=J_{1}-2J_{3}$ and $\widetilde{J}_{2}=J_{2}-J_{3}$.
Therefore, the regions of negative (i.e.~ferromagnetic) nearest and/or
next-nearest neighbor coupling become accessible.  The phase diagram of
the Heisenberg spin chain (\ref{eq-H_12}) with both positive and negative couplings is
well studied \cite{Majumdar,Haldane,Eggert,White,Hamada,Allen,Itoi}. In
addition to the antiferromagnetic and dimer phases discussed earlier, a
ferromagnetic phase exists for
$\widetilde{J}_{1}<\max\{0,-4\widetilde{J}_{2}\}$ \cite{Hamada}. The phase
diagram in terms of the effective exchange constants $\widetilde{J}_{1}$
and $\widetilde{J}_{2}$ is shown in Fig.~3.  The solid line represents
schematically the path followed in phase space, according to our numerical
calculation of the exchange constants, as the density $\nu$ increases. At
low densities, the system is close to one-dimensional and is, therefore,
antiferromagnetic. In the range of densities corresponding to an
``approximately equilateral'' configuration, the three-particle ring
exchange is strong, leading to a ferromagnetic ground state. Finally, at
even higher densities, frustration caused by the next-nearest neighbor
coupling $J_{2}$ drives the system into a dimerized phase. (Note that
there is some controversy concerning the physics of the parameter regime
$-4\widetilde{J}_{2}<\widetilde{J}_{1}<0$, where the existence of a
spectral gap associated with dimerization has not yet been established
conclusively \cite{Itoi}.)

It turns out that the above picture, based on the calculation of the
exponents to first approximation, is incomplete: because only the exchanging
particles were allowed to move while all
spectators were frozen in place, the values of $\eta_{l}$ were
overestimated. Surprisingly, allowing spectators to move results
not only in quantitative but in qualitative changes as seen in
Fig.~4. At large densities, the four-particle ring exchange $J_{4}$
dominates over $J_{2}$.  Contrary to $J_{3}$, the four-particle ring
exchange not only modifies the nearest and next-nearest neighbor
exchange constants---in addition, it introduces more complicated spin
interactions \cite{Thouless}.
For the zig-zag chain, we find
\begin{eqnarray}
H_4&\!=\!&J_4\sum_j\Big(\sum_{l=1}^3\frac{4-l}{2}
{\bf S}_j{\bf S}_{j\!+\!l}+2\big[({\bf S}_j{\bf S}_{j\!+\!1})({\bf S}_{j\!+\!2}
{\bf S}_{j\!+\!3})\nonumber\\
&&+({\bf S}_j{\bf S}_{j\!+\!2})({\bf S}_{j\!+\!1}
{\bf S}_{j\!+\!3})-({\bf S}_j{\bf S}_{j\!+\!3})({\bf S}_{j\!+\!1}
{\bf S}_{j\!+\!2})\big]\Big).
\label{eq-H4}
\end{eqnarray}
Not much is known about the physics of zig-zag spin chains with interactions of
this type. Preliminary numerical studies indicate that the ground state has zero
magnetization \cite{KMM-next}. Further work is required to identify the possibly
novel spin structures. We would also like to point out that a confining potential
of different shape might alter the outcome of the competition between the very
close values of $\eta_4$ and $\eta_2$ at high densities.

\begin{figure}[!t]
\centerline{\includegraphics[height=5.8cm,clip]{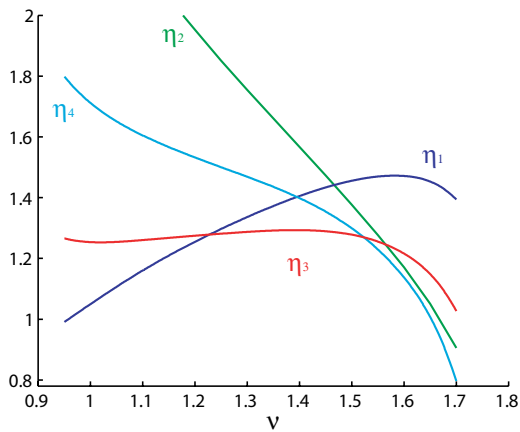}}
\caption{The exponents $\eta_{1}$, $\eta_{2}$, $\eta_{3}$, and $\eta_{4}$
as functions of the dimensionless density $\nu$. The computation
includes 12 moving spectator particles on either side of the
exchanging particles. Corrections to $\eta_{l}$ from the
remaining spectators do not exceed $0.1\%$.}
\label{fig_4}
\end{figure}

In experiments with quantum wires, the interaction strength is not a
tunable parameter: it is determined by the electron charge $e$ and the
dielectric constant $\epsilon$ in the semiconductor host. However, the
parameter $r_\Omega$ can still be tuned by adjusting the
confining potential. As $r_\Omega\propto\Omega^{-2/3}$, making the
confining potential more shallow effectively increases interaction
effects. Quantum wires in semiconductor heterostructures are fabricated
using either cleaved-edge-overgrowth or split-gate techniques. In
cleaved-edge-overgrowth wires \cite{Auslaender}, we estimate that $r_{\Omega}$
is at most of order unity due to the steep confining potential. A more shallow
confining potential is achieved in split-gate wires
\cite{Thomas,Kane,Thomas_2,Reilly,Morimoto}. Using the device specifications
of Ref.~\cite{Thomas_2}, one obtains values of $r_\Omega$ in the range
$r_{\Omega}\approx3-6$. It is not clear whether these values are large
enough to result in spontaneous spin polarization. The ideal devices for
observation of ferromagnetism would be ultra-clean wires with widely
separated gates to provide the most shallow confining potential possible.

In conclusion, interactions lead to deviations from one-dimensionality in
realistic quantum wires and, as a consequence, the Lieb-Mattis theorem no
longer applies. We have shown that strong enough interactions induce a
ferromagnetic ground state in a certain range of electron densities, where
the electrons form a zig-zag Wigner crystal.

This work was supported by the U. S. Department of Energy, Office of
Science, under Contract No. W-31-109-ENG-38.

\end{document}